\documentclass[12pt]{article}
\usepackage{amsfonts}
\usepackage{amsmath}
\usepackage[utf8]{inputenc}
\begin{document}
\newtheorem{theorem}{Theorem}
\newtheorem{lemma}{Lemma}
\newtheorem{example}{Example}
\newtheorem{definition}{Definition}
\newtheorem{corollary}{Corollary}
\newtheorem{proposition}{Proposition}
\newcommand{\ord}{\mathop{\mathrm{ord}}\nolimits}
\newcommand{\Ord}{\mathop{\mathrm{Ord}}\nolimits}
\newcommand{\supp}{\mathop{\mathrm{supp}}\nolimits}
\newcommand{\es}{\mathop{\mathrm{ess\,sup}}\nolimits}
\newcommand{\Id}{\mathop{\mathrm{Id}}\nolimits}
\newcommand{\Tr}{\mathop{\mathrm{Tr}}\nolimits}
\newcommand{\id}{\mathop{\mathrm{Id}}\nolimits}
\newcommand{\Lip}{\mathop{\mathrm{Lip}}\nolimits}
\newcommand{\Qp}{\mathop{\mathbb Q_p}\nolimits}
\newcommand{\Zp}{\mathop{\mathbb Z_p}\nolimits}

\righthyphenmin=2
\title{Entropy gain in $p$-adic quantum channels}  
\author{Evgeny Zelenov}
\date{October 3, 2020}
\maketitle

\begin{abstract}
The construction of  the $p$-adic quantum Gaussian channel is suggested. Entropy gain is calculated. The adelic formula for the entropy gain is obtained.
\end{abstract}

\label{sec:intro}
\section{Introduction}
Starting from \cite{V, V1}, non-Archimedean analysis has been actively used to build physical models. Thereby, the branch of mathematical physics (p-adic mathematical physics) has arisen. One can find a bibliography on this subject in the book \cite{VVZ} and reviews \cite{DKhKV, DKhKVZ}.

The article is organized as follows. Firstly, we give the necessary facts of the p-adic analysis. Secondly,
we define the p-adic Gaussian state and the $p$-adic Gaussian linear bosonic channel and also give their properties. This information follows \cite{Chan}; the results are presented without proof. New results are shown below. Thirdly, we prove a formula for the magnitude of the entropy gain in the $p$-adic Gaussian channel. Fourthly, we give the adelic formula for the entropy gain and its possible applications.

\label{sec:p-adics}
\section{$p$-Adic numbers and symplectic geometry}	
In this section, some known facts concerning the geometry of lattices in a two-dimensional symplectic space over the field $\Qp$ of $p$-adic numbers are presented without proof. The necessary information from the $p$-adic analysis is contained, for example, in \cite{Sch}. Most statements regarding the geometry of lattices can be found in \cite{S}.

Let $F$ be a two-dimensional vector space over the
field	$\Qp $.	A nondegenerate	 symplectic	form
$\Delta\,\colon\,F\times F\,\to\,\mathbb Q_p$ is given in the space $F$. A free module of rank two over the ring $\Zp$ of $p$-adic integers considered as a subset of $F$  will be called a lattice in $F$. A lattice is a compact set in the natural topology in the space $F $.
We introduce the duality relation on the set of lattices. Let $L$ be a lattice; then the dual lattice $L^*$ means the following subset of the space $F$:
$u\in L^*$ if and only if the condition $\Delta (u,v)\in\mathbb Z_p$ is satisfied for all $v\in L$.
If $L$ coincides with $L^*$, then the lattice $L$ will be called a self-dual lattice.
For any self-dual lattice $L$, there exists a symplectic 
basis $\left(e_1, e_2\right)$ in the space $F$
such that the lattice $L$ has the form
$$
L=\mathbb Z_p e_1\bigoplus\mathbb Z_p e_2.
$$
By $Sp (F,\Delta)$ denote the symplectic group (the
group of nondegenerate linear transformations of the space $F$ that preserve the form $\Delta$). The group $Sp (F,\Delta)$ is isomorphic to the group $SL_2\left(\mathbb Q_p\right)$.
In the space $F$, there exists a unique translation-invariant measure (Haar measure) up to normalization. We will normalize the measure in such a way that the measure of a self-dual lattice is equal to unity. The action of the symplectic group preserves the measure; therefore, the measure of any self-dual lattice is equal to unity. The measure of the lattice $L$ will be denoted
by $|L |$. If $L$ is a self-dual lattice, then, as noted above, $|L|=1$; the converse is also true, if $|L|=1$, then the lattice $L$ is self-dual. It is easy to verify the validity of the relation  $|L||L^*|=1$.
Let $S\in Sp (F,\Delta)$, $L\subset F$ be an arbitrary lattice. As already noted, the action of the symplectic group preserves measure, that is, $|SL|=|L|$ . The converse is also
true, if  $L_1,\, L_2$  are arbitrary lattices in  $F$  with the same
measure, $|L_1|=|L_2|$ ,  then  there  exists  a  symplectic
transformation $S\in Sp (F,\Delta)$,  such that $SL_1=L_2$.

\label{sec:channels}
\section{$p$-Adic linear bosonic channels}

Let $\mathcal H$ be a separable Hilbert space over the field $\mathbb C$ of complex numbers. The scalar product in $\mathcal H$ will be
denoted by  $\langle\cdot, \cdot \rangle$, and we will consider it antilinear in
the first argument.
The state of the system is described by the density
matrix $\rho$ in the space $\mathcal H$. Denote by $\mathfrak S$ the set of all states. By
$\mathcal B (F)$  we denote the   $\sigma$-algebra of Borel subsets
of  $F$;  by  $\mathfrak B (\mathcal H)$ we denote the algebra of
bounded operators on space $\mathcal H$. A quantum observable $M:\mathcal B (F) \to \mathfrak B (\mathcal H)$ is a projection-valued measure on $\mathcal B (F)$. Denote by $\mathfrak M$ the set of quantum
observables.
The Born–von Neumann formula gives the probability distribution of the observable $M$ in a state $\rho$
$$
\mu_\rho^M = \mathrm{Tr}\left(\rho M(B)\right), \,B\in\mathcal B (F).
$$
 
We do not go beyond the framework of the standard statistical model of quantum mechanics (\cite{H1}), since the set of real numbers and the set of $p$-adic numbers are Borel isomorphic.

Let $W$ be a mapping from the space $F$ to the set of
unitary operators in the space $\mathcal H$, satisfying the relation
$$
W(z)W(z') = \chi\left(\frac{1}{2}\Delta (z,z')\right)W(z+z')
$$
for all $z,z'\in F$. Here we use the notation $\chi (x) = \exp\left(2\pi i\{x\}_p\right)$, where $\{x\}_p$
denotes the p-adic fractional part of the number of $x\in\mathbb Q_p$.
Besides, we assume that mapping $W$ is continuous
in the strong operator topology.  The mapping $W $ is
called the representation of canonical commutation relations (CCRs) in the Weyl form.
As noted above, the state of a quantum system is described by a density matrix $\rho$ in a Hilbert space $\mathcal H$. Let   an   irreducible   representation   of   the  CCR
$\{W(z),\,z\in F\}$ be given in this space. Each density
operator can be associated with a function $\pi_\rho$ in the space $F $ by the following formula.
$$
\pi_\rho (z) = {\mathrm Tr}\left(\rho W(z)\right).
$$
The function $\pi_\rho$ is called the characteristic function of the quantum state $\rho$ and defines this state
uniquely. The characteristic
function is used to reconstruct the state $\rho$ according to the relation
$$
\rho = \int_F\pi_\rho(z)W(-z)dz.
$$ 
The characteristic function of a quantum state has
the property of $\Delta$-positive definiteness: for any finite
sets $z_1, z_2, \dots , z_n$  of points in the space $F$ and complex numbers $c_1, c_2, \dots , c_n$, the inequality
$$
\sum_{i,j=1}^n c_i\bar c_j\pi_\rho (z_i-z_j)\chi\left(-\frac{1}{2}\Delta (z_i,z_j)\right)\geq 0
$$
is satisfied.
As in the case of the representation of CCR over a
real symplectic space, a noncommutative analogue of
the Bochner–Khinchin theorem holds for the $p$-adic
case; this theorem establishes  one-to-one correspondence
between states and $\Delta$-positive definite
functions (\cite{BKH})
.
We give the following definition.

\begin{definition}
A state $\rho$ will be called a  $p$-adic
Gaussian state if its characteristic function is the
 indicator function of some lattice, that is,
$$
\pi_\rho(z) = {\mathrm Tr}\left(\rho W(z)\right) = h_L(z) = \begin{cases}
1, &\text{ $z\in L$;}\\
0, &\text{  $z\notin L$.}
\end{cases}
$$
\end{definition}
This definition is natural in the following context.
Let $\mathcal F$ be the Fourier transform in $L^2(F)$ defined by
the formula
$$
\mathcal F \left[f\right](z) = \int_F\chi\left(\Delta (z,s)\right)f(s)ds,
$$
$L$ is a lattice in $F$. Then, the formula
$$
|L|^{-1/2}\mathcal F\left[h_L\right] = |L^*|^{-1/2}h_{L^*}
$$
is satisfied.
In other words, the Fourier transform turns
the characteristic function of the lattice into the characteristic
function of the dual lattice up to a factor. In
particular, the characteristic function of the self-dual
lattice is invariant under the action of the Fourier
transform. In this context, the characteristic function
of the lattice is an analogue of the Gaussian function in
real analysis.

$p$-Adic Gaussian states are elementary in structure.
Namely, the following statements are true.
\begin{proposition}
\label{gs}
The characteristic function $h_L$
of the lattice  $L$ determines the quantum state,
if and only if the condition $|L|\leq 1$
is satisfied.

A Gaussian state $\rho$ having a characteristic function $\pi_\rho = h_L$
is  $|L|P_L$, where $P_L$ is an orthogonal projector of
rank $1/|L|$.
\end{proposition}
Some obvious properties of Gaussian states are
given below.
\begin{proposition}
\label{propgauss}
The following statements are true.
\begin{itemize}
\item
A Gaussian state is pure if and only if the corresponding
lattice is self-dual.
\item
The entropy of the Gaussian state defined by the
lattice $L$ is $-\log |L|$.
\item
Gaussian states $\rho_1$ and $\rho_2$ are unitarily equivalent
if and only if the corresponding lattices $L_1$ and $L_2$ have
the same measure.
\item
The entropy of the Gaussian state determines
this state uniquely up to unitary equivalence.
\item
The Gaussian state has the maximum entropy
among all states with a fixed rank $p^m,\,m\in\mathbb N$.
\end{itemize}
\end{proposition}

We will use the notation $\gamma (L)$ for the density operator
of the Gaussian state defined by the lattice $L$. We
considered only centred Gaussian states. We can similarly
view general Gaussian states $\gamma (L,\alpha )$, which
are defined by a characteristic function of the form
$$
\pi_{\gamma (L,\alpha )} = \chi\left(\Delta (\alpha, z)\right)h_L(z).
$$
It is easy to see that
$$
\gamma (L,\alpha ) = W(\alpha )\gamma (L) W(-\alpha ).
$$
Let $W$ be the irreducible representation of the
CCR in the Hilbert space $\mathcal H$; $\mathfrak S (\mathcal H)$ is the set of states.

By analogy with the real case (\cite{H2}), a linear
bosonic channel (in the Schrodinger representation) is
a linear completely positive trace-preserving mapping $\Phi :\mathfrak S (\mathcal H)\to \mathfrak S (\mathcal H)$
such that the characteristic function  $\pi_\rho$
of any state $\rho\in \mathfrak S (\mathcal H)$ is transformed by the
formula
\begin{equation}
\label{chanel}
\pi_{\Phi [\rho ]}(z) = \pi_\rho (Kz)k(z)
\end{equation}
for some linear transformation $K$ of the space $F$ and
some complex-valued function $k$ in $F$.

Generally speaking, expression (\ref{chanel}) does not always
determine the channel; to do this, additional conditions
on the transformation $K$ and the function $k$ are
necessary.

A $p$-adic Gaussian channel is a linear bosonic
channel for which the function $k$ is the characteristic
function of some lattice $L\subset F$, that is, $k(z) = h_L(z), z\in F$.

The following proposition holds.
\begin{proposition}
\label{gqc}
Let $K$ be a nondegenerate linear
transformation of the space $F$, $L$ be a lattice in the
space $F$, $k(z) = h_L(z)$. In this case, expression (\ref{chanel})
defines a channel if and only if the inequality
\begin{equation}
\label{ineq}
|1-{\mathrm det} K|_p|L|\leq 1
\end{equation}
\end{proposition}
is satisfied.
Note that in the case of ${\mathrm det} K = 1$, the transformation
is symplectic and, therefore, is unitarily representable.
Next, we consider the case of ${\mathrm det} K \neq 1$.

\label{sec:entropy}
\section{Entropy gain} 
The entropy $H(\rho)$ of the state $\rho$ is defined by the following
expression $$H(\rho) = -\Tr\rho\log\rho.$$

According to the definition, the entropy gain is
$$
G(\Phi ) = \inf_\rho\left\{H\left(\Phi [\rho ]\right) - H(\rho )\right\}.
$$
\begin{theorem}
The following formula is valid for the
entropy gain of the $p$-adic Gaussian linear bosonic  channel
$$G(\Phi ) = \log |\det K|_p.$$
\end{theorem}
Let $K$ be a nondegenerate linear transformation
and $L$ be a lattice. Moreover, we assume that $K$ and $L$
define the channel, that is, they satisfy condition (\ref{ineq}).
By $L_n,\,n\in\mathbb N$ we denote the lattice $L_n=p^nL$. Note that $L_n\subset L$, and for sufficiently large $n$, the condition $|L_n|<1$
is satisfied, since $|L_n|=p^{-n}|L|$. According to
Proposition \ref{gs}, the Gaussian state $\gamma (L_n)$ is defined for
all such $n$. We also note that there exists such $N\in\mathbb N$,
that for all $n\geq N$, the conditions  $L_n\subset L$ and $K^{-1}L_n\subset L$
are simultaneously satisfied. It is not difficult
to calculate the entropy gain for state $\gamma (L_n),\,n\geq N$.
Indeed, the condition
$$
\pi_{\Phi\left[\gamma (L_n)\right]}(z) = \pi_{\gamma (L_n)}(Kz)h_L(z) = h_{L_n}(Kz)h_L(z) = h_{K^{-1}p^nL}(z)
$$
follows directly from the definition of the Gaussian channel
(\ref{chanel}).
Consequently,
\begin{equation}
\label{fgs}
\Phi\left[\gamma (L_n)\right] = \gamma\left(K^{-1}L_n\right)
\end{equation}
and, in accordance with Proposition \ref{gs},
\begin{equation}
\label{eg}
H\left(\Phi\left[\gamma (L_n)\right] \right) - H\left(\gamma(L_n)\right) = \log|\det K|_p.
\end{equation}
The latter formula immediately implies an evaluation
for the minimum entropy gain
$G(\Phi)\leq \log|\det K|_p$.
Further, we recall (Proposition \ref{gs}) that operator $|L_n|^{-1}\gamma(L_n)$
is an orthogonal projection, and the
sequence of these operators strongly converges to the
unit operator  $I$ when $n\to\infty$. Therefore, we can correctly
determine $\Phi[I]$ (the channel is regular); moreover,
formula (\ref{fgs}) and Proposition \ref{gs} imply $\|\Phi[I]\| = |\det K|_p^{-1}$. Next, we use the evaluation of the
minimum entropy gain for a regular channel $-\log\|\Phi[I]\|\leq G(\Phi)$ obtained
in \cite{H3}. The theorem is proved.
Note that the expression for the entropy gain in the
$p$-adic linear bosonic Gaussian channel is given by a
a formula similar to the corresponding expression
for the real linear bosonic Gaussian channel obtained
in \cite{H3}.

\section{Adelic channels and entropy gain}

Now let $F$ be a two-dimensional vector space over
the field $\mathbb Q$ of rational numbers, $\Delta$ be a nondegenerate
symplectic form taking values in the field $\mathbb Q$, and $K$
be a nondegenerate linear transformation of the
space $F$.
For each prime $p$, we construct the corresponding
linear bosonic Gaussian channel $\Phi_p$ and calculate the
entropy gain $G(\Phi_p)$ of such a channel. We can also
construct a real linear bosonic Gaussian channel $\Phi_\infty$
and calculate the entropy gain $G(\Phi_\infty)$. Let $\mathcal P$ denotes the set of all prime numbers. The following
statement is true.
\begin{theorem}
\label{adele}
$$
G(\Phi_\infty) + \sum _{p \in\mathcal P}G(\Phi_p) = 0.
$$
\end{theorem}
Note that $\det K\in\mathbb Q$. The further follows from a
simple adelic formula
$$
|\det K|\prod_{p \in\mathcal P}|\det K|_p = 1,
$$ 
which is valid for an arbitrary
nonzero rational number (for example, see \cite{Sch}).

Theorem \ref{adele} can be interpreted as follows. Consider
the adelic linear bosonic Gaussian channel generated
by the linear transformation $K$ of a two-dimensional
vector space over the field of rational numbers, that is,
the tensor product of the channels $\Phi_p$  over all primes $p$
and the real channel $\Phi_\infty$. A nontrivial entropy gain $G(\Phi_p)$
is possible in each component of this adelic
channel. However, the total entropy gain in the adelic
channel is zero. There is a nontrivial exchange of
information between the components of the adelic
channel when the total entropy is conserved.
If we accept the hypothesis that <<at the fundamental
level, our world $\dots$ is adelic $\dots$>>, then Theorem \ref{adele}
can be used, for example, to interpret the black hole information
paradox.

\end{document}